\documentclass[11pt]{article}
\usepackage{moriond,epsfig}

\def\be{\begin{equation}}
\def\ee{\end{equation}}
\def\bea{\begin{eqnarray}}
\def\eea{\end{eqnarray}}

\begin{document}
\vspace*{4cm}
\title{LITTLE HIGGS SEARCHES AT LHC}

\author{ J.E. GARCIA }

\address{IFIC, University of Valencia - CSIC, \\
46071 Valencia, Spain
\vskip 0.5cm
\large{On behalf the ATLAS collaboration}}

\maketitle\abstracts{ 
A new method of solving the hierarchy problem in the SM has been proposed.  This method leads to the so-called  "Little Higgs" models. The ATLAS experiment at LHC has undertaken studies of the new particles predicted by these model: a heavy top quark, heavy gauge bosons and additional Higgs bosons. Simulations of their decays have been carried out. The sensitivity of the ATLAS experiment to discover these new particles is discussed.}

\section{Little Higgs Models}
\label{sec:lhiggs}

\subsection{Introduction}
\label{sec:intro_lhiggs}

The Higgs sector of the Standard Model (SM) contains an instability due to the radiative corrections to the Higgs boson mass. A fine tuning is required to have a Higgs boson with a mass of the order of 100 GeV. This problem is known as the hierarchy problem of the SM.  Various theories may solve the problem in different ways. Supersymmetry~\cite{susy1,susy2} stabilizes divergent loop contributions to the Higgs mass introducing new fermions and bosons. Novel theories with extra dimensions exploit the geometry of a higher dimensional space-time to resolve the hierarchy problem~\cite{ed}.

Recently a new class of theories with the desired cancellation of quantum corrections was discovered. This scenario is known as \emph{Little Higgs}~\cite{lhiggs_first}. The idea is to introduce the Higgs as a pseudo-Goldstone boson resulting from a spontaneously broken global symmetry at a scale $\Lambda_S$ of the order of 10 TeV. In this way, the Higgs boson remains light, protected by this approximate global symmetry and the cancellation of 1-loop quadratic corrections.

Several implementations of this model exist, described~\cite{models,littlest,model_2} by the global symmetry breaking group, $SU(5)/SO(5)$, $SU(6)/SP(6)$, the minimal moose~\cite{min_moose}  $SU(3)^2/SU(3)$ and general mooses~\cite{gen_moose} $SU(3)^n/SU(3)^k$. The UV cut-off of these theories is $\Lambda_S \sim 10~TeV$. Above the scale $\Lambda_S$, Little Higgs theories break down.

All Little Higgs Models have necessarily in common a set of new particles. One heavy top-like particle with $2/3$ electric charge is required to cancel the contribution to $m_H$ of the top quark. A gauge boson sector is needed to cancel the loops of SM gauge bosons. A scalar sector is required as well to cancel corrections from the Higgs boson itself. The masses of all these new particles is of the order of 1 TeV. 

\begin{figure}[hpbt]
  \begin{center}
    \psfig{figure=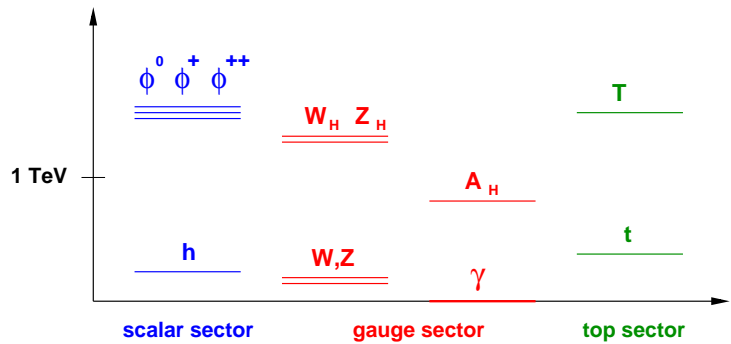,width=0.6\linewidth}
   \end{center}
\caption{Particles predicted in the Littlest Higgs models, with corresponding masses.\label{fig:masses}}
\end{figure}

The minimal model containing the Little Higgs ideas is called \emph{Littlest Higgs} model~\cite{littlest}. This model is based on a non-linear sigma model including the $SU(5)$ to $SO(5)$ symmetry breaking. It contains a $SU(5)$ global symmetry with two copies of the electroweak symmetry group, $[SU(2) \otimes U(1)]^2$. At the scale $\Lambda_S$, the global symmetry $SU(5)$ is spontaneously broken to $SO(5)$ via a \emph{vacuum expectation value} of order $f$ (scale parameter with value around 1 TeV). The symmetry scale $\Lambda_S$ takes the value $\Lambda_S \sim 4 \pi f$. At the same time $[SU(2) \otimes U(1)]^2$ is also broken into the electroweak group,  $SU(2)_L \otimes U(1)_Y$. The broken global symmetry yields 14 massless Goldstone bosons. These massless bosons acquire mass, due to the broken symmetries, at the scale $f$. Heavy partners of the electroweak gauge bosons $A_H,~W_H^\pm$ and $Z_H$ and scalar particles $\phi^0,~\phi^+,~\phi^{++}$ appear, in addition to a new $2/3$ charge heavy quark (T). In this breaking process, the Higgs mass is preserved with a low value within the range expected according to LEP experimental data. In the following the Standard Model Higgs will be denoted $h$. Figure~\ref{fig:masses} shows the particle mass spectrum predicted by the model~\cite{phenomenology}.

The searches for these new particles have been performed using ATLAS Monte Carlo generators and analysis tools. For the simulation of the various processes, the \textsc{Pythia}~\cite{pythia} generator has been used. Couplings and masses were modified according to reference~\cite{phenomenology} to match the model parameters. The events were passed through the ATLAS detector simulation program ATLFAST~\cite{atlfast}. This package gives a parametrized detector response based on a more detailed simulation of the ATLAS detector using the GEANT program~\cite{geant}. The output of the detector simulation was used for the analysis. The cuts performed fulfill the ATLAS trigger requirements~\cite{hlt} and detector capabilities~\cite{atlas}.

 The searches using simulations of the ATLAS detector, production cross sections and the decay channels for the new particles are reviewed in the following sections.

\subsection{Heavy top}
\label{sec:T}

The loop corrections to the Higgs mass coming from the SM top is the largest contribution. In the SM, fermions acquire their mass according to the Higgs mechanism via Yukawa interactions. In the same way, in the Little Higgs model two Yukawa couplings are needed, $\lambda_1$ and $\lambda_2$. The mass and the Yukawa coupling of top quark are fixed in the SM. In order to recover these SM values, $\lambda_1$ and $\lambda_2$, should satisfy some constraints. These requirements allow the cancellation of SM top-quark corrections. Equation~\ref{eq:t_mass} shows the value of the $T$ quark mass as function of the Yukawa couplings and the condition to be satisfied by these couplings in order to be compatible with the SM values,


\begin{equation}
M_T = f \sqrt{\lambda_1^2 + \lambda_2^2} \qquad \frac{1}{\lambda_1^2} + \frac{1}{\lambda_2^2} = \Big (\frac{v}{m_t} \Big )^2 \approx 2  
\label{eq:t_mass}
\end{equation}

\noindent
where $m_t$ is the top mass and $v$ is the so-called \emph{Fermi scale}, $v = 244~GeV$. The mass $M_T$ is of the order of the scale $f$. The constraint that none of the loop contributions exceeds the value of the Higgs mass squared by more than a factor of 10 ($\sim 10$\% fine-tuning) yields an upper bound on $M_T$ as a function of $m_H$. Lower bounds can also be extracted from Equation~\ref{eq:t_mass}. Therefore $M_T$ should be inside the mass window:

\begin{equation}
\frac{2 m_t}{v} f ~\leq ~M_{T} ~\leq ~2~TeV \Big( \frac{m_H}{200~GeV} \Big)^2
\label{eq:t_bounds}
\end{equation}

\begin{figure}[tbp]
\begin{center} 
  \psfig{figure=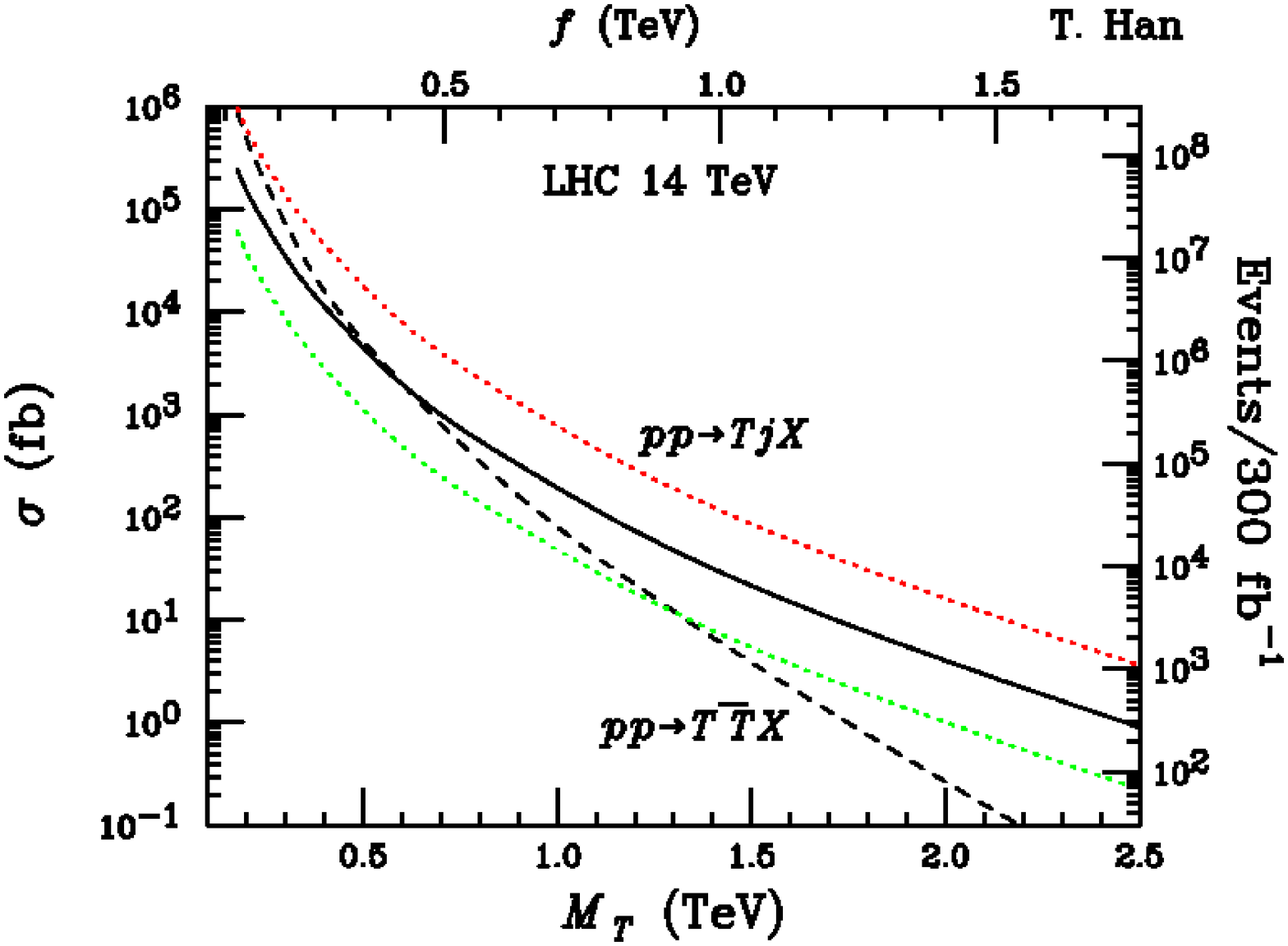,width=0.48\linewidth,height=6.2cm}
  \psfig{figure=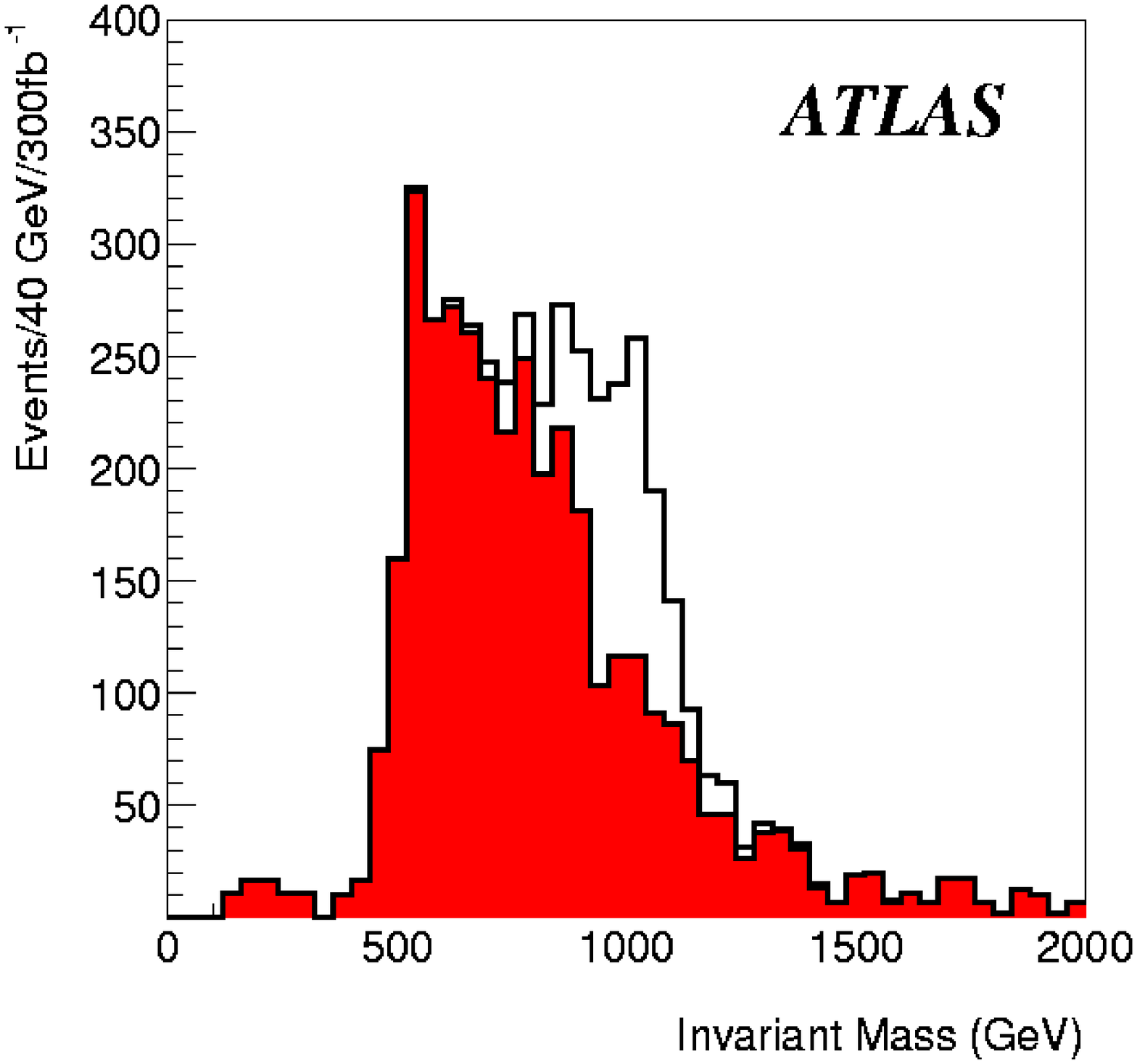,width=0.38\linewidth}
\end{center}
\caption{Cross section for the new top-like quark, $T$, versus mass (left) and the invariant mass reconstruction signal and background (dark) of the $T$ quark using the $Wb$ channel in the ATLAS detector for an integrated luminosity of 300 $fb^{-1}$ (right).\label{fig:new_part}}
\end{figure}

\noindent
where $m_H$ is the expected mass of the Higgs boson. Figure~\ref{fig:new_part} (left) from reference~\cite{lhiggs_sn} shows the production cross section of $T$ as function of $M_T$ at LHC energies ($\sqrt{s} = 14~TeV$). $T \overline T$ production (dashed line) and single $T$ production (solid and dotted) are shown. The solid line corresponds to a choice $\lambda_1 / \lambda_2 = 1$, whereas the dotted line corresponds to $\lambda_1/\lambda_2 = 2$ and $1/2$. The $Wb$ fusion mechanism is dominant for masses larger than 700 GeV. $T \overline{T}$ production dominates for lower masses. In this analysis, the mass of $T$ is considered to be above 700 GeV, thus the production via $Wb$ fusion has been used. $T$ has 3 possible decay modes, with the following partial widths:

\begin{equation}
\Gamma(T \to t Z) = \Gamma(T \to t~h) = \frac{1}{2} \Gamma(T \to b W) = \frac{\kappa_{T}^{2}}{32 \pi} M_T
\label{eq:top_br}
\end{equation}

\noindent
where $\kappa_T = \lambda_{1}^{2}/\sqrt{\lambda_{1}^{2}+\lambda_{2}^{2}}$. Other decay channels are suppressed by a factor $v^2/f^2$. The first two decays are distinctive of Littlest Higgs models.

These three decay channels have been studied. The more advantageous signature associated with the $Zt$ decay is: 3 isolated leptons (2 of them from the $Z$ decay), one $b$-jet and missing energy. The following requirements are applied: 3 high $p_T$ leptons, one $b$-tag and large $E_T(miss)$. The background is largely dominated by $WZ$ production, although other processes like $t \overline t$ and $tbZ$ have significant contributions. This channel has a low number of expected events (around 10) for a $T$ mass of 1 TeV and an integrated luminosity of 300 $fb^{-1}$ (3 years of LHC operation at high luminosity). 

The decay of the quark $T$ to $ht$ is also difficult to observe. The topology of the event depends on the Higgs mass. In all these analyzes a Higgs mass of 120 GeV has been assumed, so the $h$ decay to $b \overline b$ dominates. The final state corresponding to the $T \to ht$ decay is an isolated lepton and 3 $b$-jets with missing energy. At least one tagged $b$-jet is required, since a more severe requirement does not reduce the background but decreases the efficiency. The background is dominated by $t \overline t$ production and only kinematics can resolve signal and background. The discovery in this mode is therefore very difficult. 

More favorable is the case of $T$ decaying to $Wb$ (see Figure~\ref{fig:new_part} right). The branching ratio of this decay is larger by a factor 2, but the background is also larger. Using the high $p_T$ of the lepton, $E_T(miss)$ from the W decay and the requirement of one tagged $b$-jet (out of the 2 $b$-jets in the event), the background can be reduced. An excess of signal over background can be observed for a mass of 1 TeV. The two main sources of background in this channel are: $Wbb$ and $t \overline t$ production. A veto on the total number of jets and a cut in the invariant mass of the two jets are critical to reject the background. 

All these analyzes have been performed assuming $\lambda_1 = \lambda_2 = 1$. A larger value of $\lambda_1 / \lambda_2$ would result in a larger signal in all previous cases.

\subsection{Heavy gauge bosons}
\label{sec:bosons}


The new boson sector includes three gauge bosons, two neutrals, $A_H$ and $Z_H$, and the charged $W^\pm_H$. $A_H$ is the massive partner of the photon $\gamma$ and is typically lighter than the other gauge bosons. This boson could be produced at LHC, but also at the Tevatron, provided the mass is light enough. The mass of $A_H$ is dependent on the parameters of the model and has in addition theoretical uncertainties related to the couplings. $A_H$ can be searched at colliders via the decays $A_H \to f \overline f$ and $A_H \to Z~h$. Due to the additional uncertainties mentioned before, the results concerning this particle are not discussed here (see reference~\cite{lhiggs_sn} for details).


The broken $SU(2)$ group yields 3 new particles $W^\pm_H$ and $Z_H$. The mass of these bosons is again of the order of $f$. The mass of the SM bosons is also affected by corrections of the order of $v^2/f^2$. The tree level SM relations $M_W/M_Z = c_W$ and $\rho = 1$ are no longer valid. The $SU(2)$ custodial symmetry of the SM is therefore broken. Precision measurements on electroweak observables, like $\rho$, can constrain the model and provide a lower bound on the scale $f$. The Littlest Higgs model may be modified in order to preserve the custodial $SU(2)$ symmetry at first order~\cite{custodial,deandrea}, but these models are not considered here. The following bounds on the masses can be derived as before:

\begin{equation}
m_W \frac{2 f}{v}  ~\leq ~M_{W_H} \approx  M_{Z_H}  ~\leq ~6~TeV \Big( \frac{m_H}{200~GeV} \Big)^2
\label{eq:gauge_bounds}
\end{equation}

\begin{figure}[h]
\begin{center}
  \psfig{figure=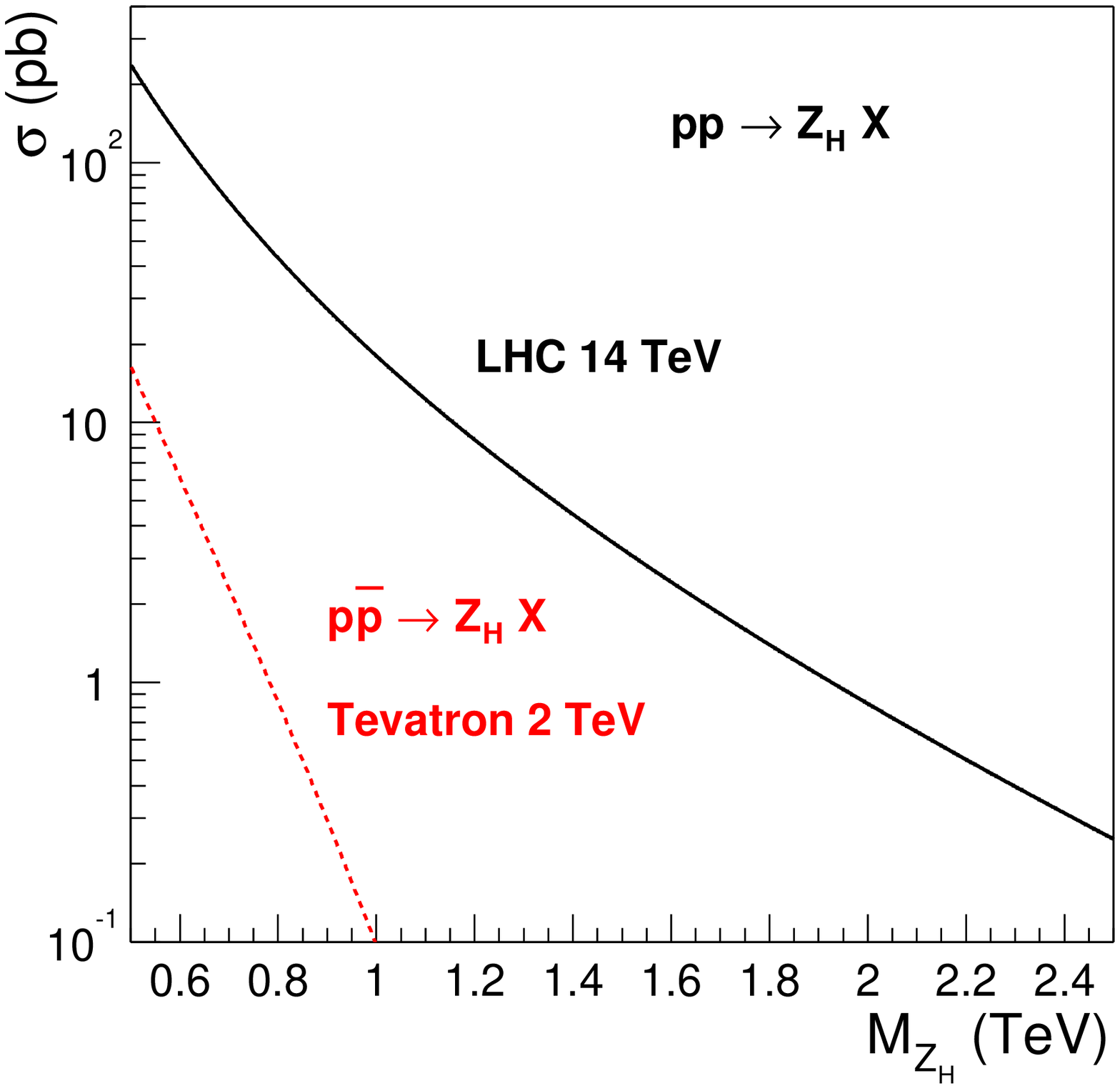,width=0.48\linewidth}
  \psfig{figure=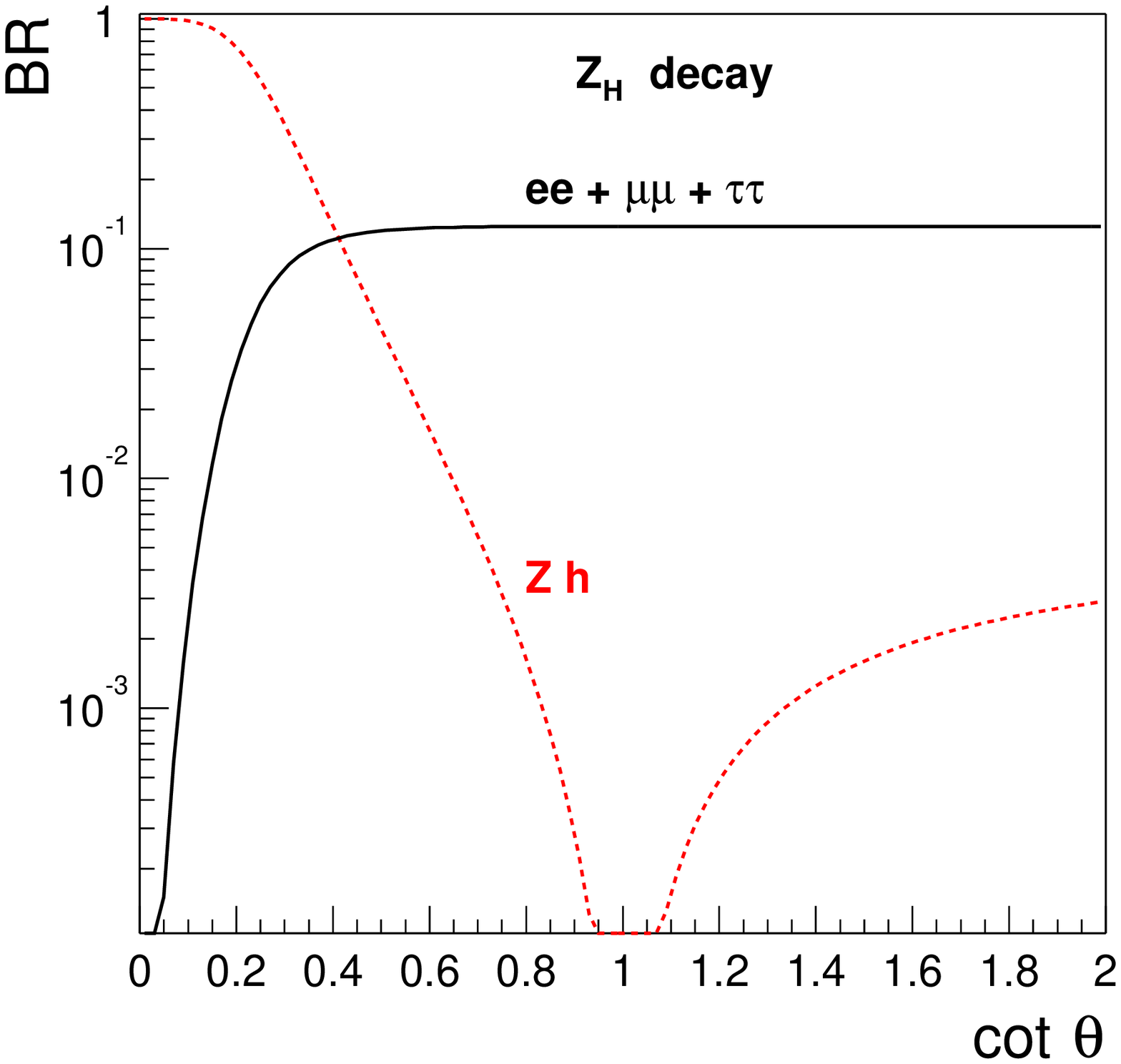,width=0.48\linewidth}
\end{center}
\caption{Production cross section for $Z_H$ gauge boson production versus $Z_H$ mass assuming $\cot{\theta} = 1$ (left) and branching ratio for the different decay channels versus $\cot{\theta}$ (right).\label{fig:zh}}
\end{figure}

Figure~\ref{fig:zh} (left) shows the cross section for $Z_H$ production versus mass at the LHC. The $W_H$ cross section is larger by a factor 2. Only one parameter besides the mass is needed to obtain the production of $Z_H$ and $W_H$, namely $\cot{\theta}$ ($\theta$ is a mixing angle analogous to the Weinberg angle). In the Figure, a value of $\cot{\theta} = 1$ has been chosen. The cross section is proportional to $\cot^2{\theta}$ since $Z_H$ and $W_H$ couplings to fermion pairs are proportional to $\cot{\theta}$. It follows that LHC will be able to produce a large number of heavy $Z_H$ particles. It also seems possible to exploit all the available decay channels to search for these heavy gauge bosons.

The branching ratios of $Z_H$ versus the parameter $\cot{\theta}$ are presented in Figure~\ref{fig:zh} (right). Two main decays are present, to fermion pairs and to a SM gauge boson in association with the Higgs boson. Neglecting $QCD$ corrections and fermion masses (also $top$ quark) the total widths of $Z_H$ and $W_H$ are equal. The width is:

\begin{equation}
 \frac{\Gamma}{M} = \big[ 3.4\cdot \cot^2 \theta+0.071\cdot \cot^2 2\theta
 \big] \%
\end{equation}

\noindent
where the first term includes all possible fermion decays and the second term only the $h$ decay mode.

In Figure~\ref{fig:zh} (right) the decays of $Z_H$ are displayed versus $\cot{\theta}$. For a value of $\cot{\theta}$ smaller than 0.5, the decay to $Z~h$ is dominant. But for $\cot{\theta} \sim 1$, and due to the proportionality of the cross section to $\cot{\theta}$, the production is suppressed. This decay is a particularly relevant channel to test the model. For $\cot{\theta} > 0.5$ the branching ratio to the 3 generations of charged leptons is 12 \%. The decay into any quark pair ($b \overline b$, $t \overline t$,...) is also equal to 12 \%.

\begin{figure}[hpbt]
  \begin{center}
    \psfig{figure=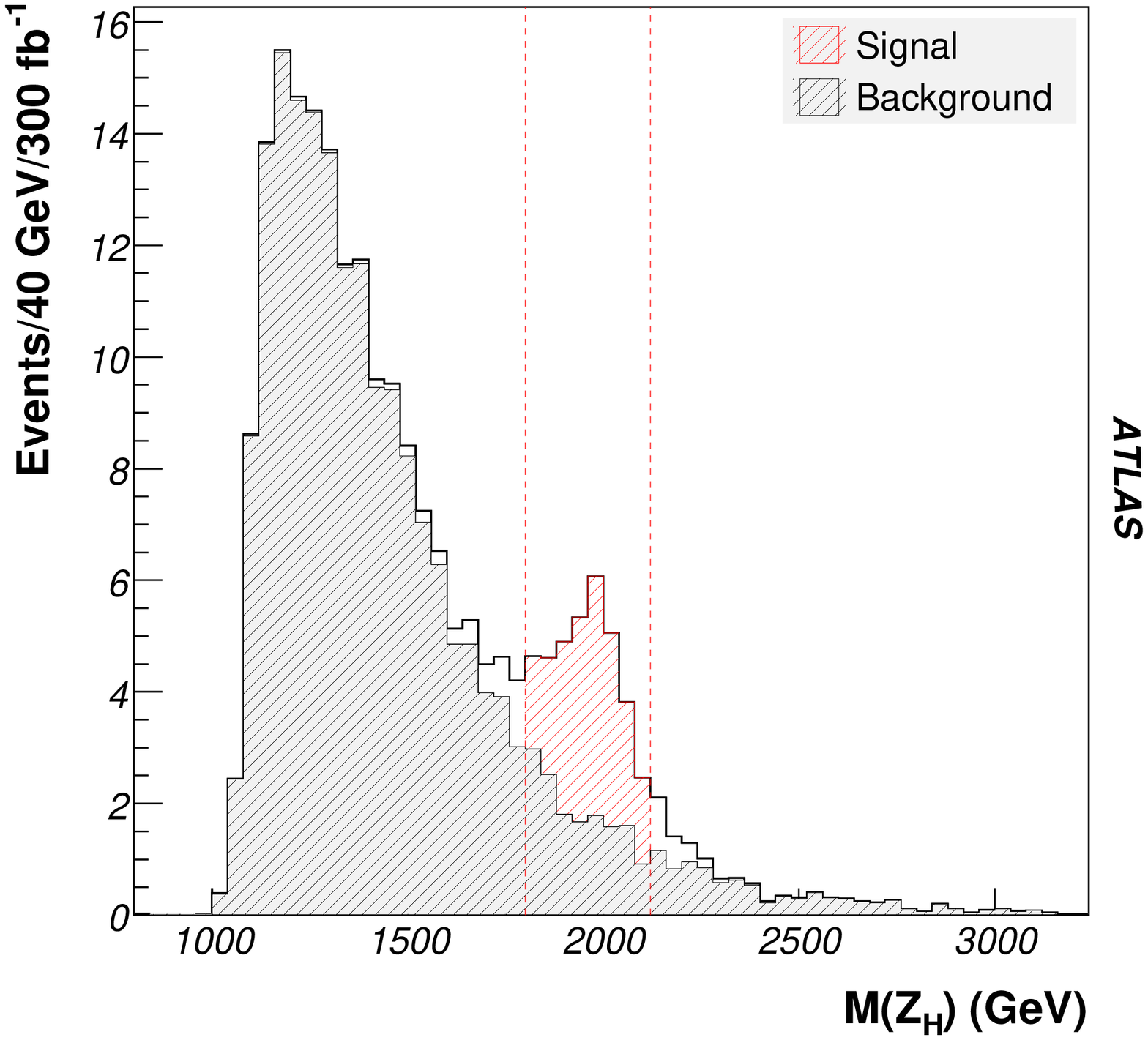,width=0.49\linewidth}
    \psfig{figure=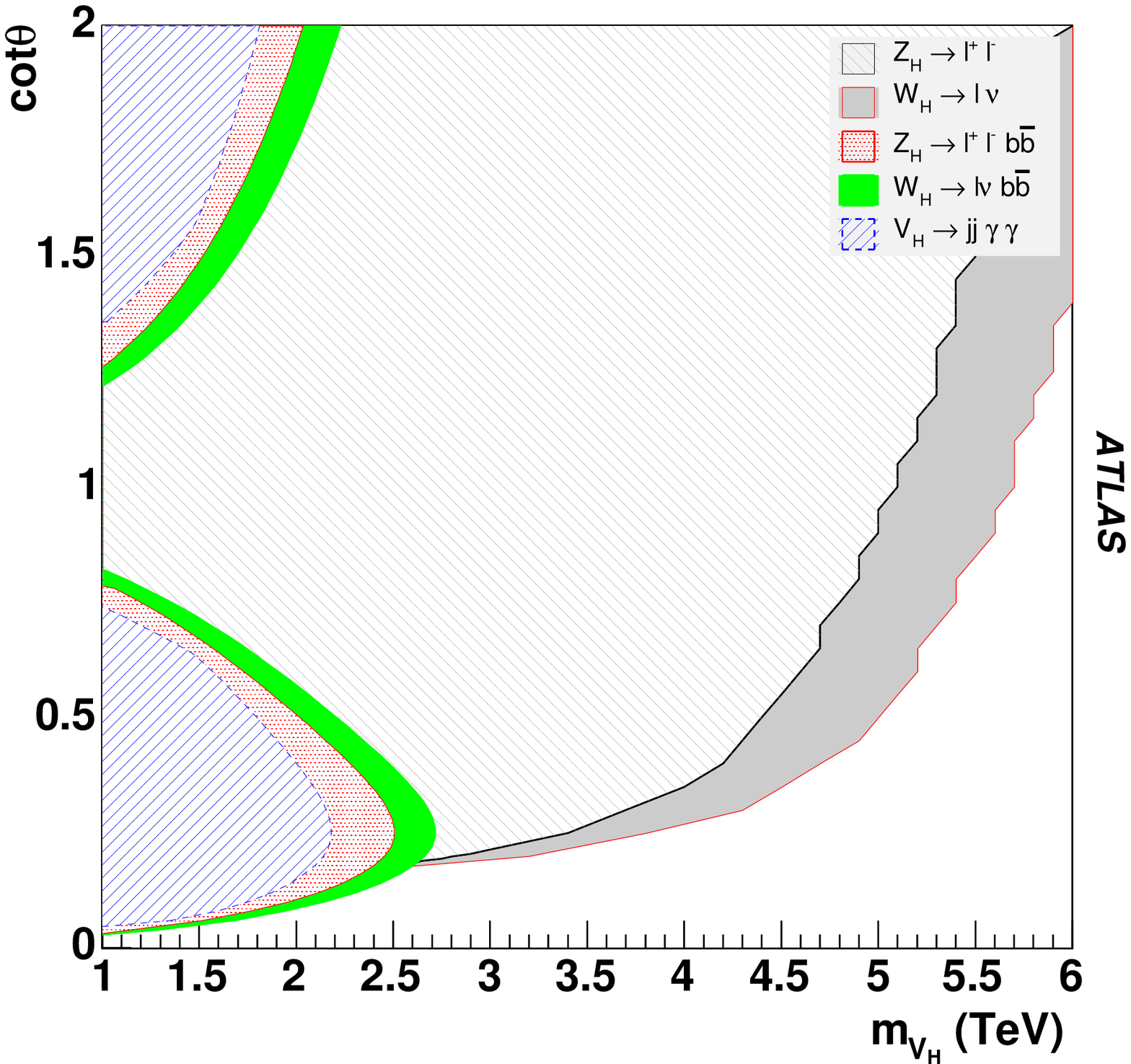,width=0.48\linewidth}
  \end{center}
  \caption{On the left, the background and the invariant mass of the $Z_H$ reconstructed using the $Zh$ channel with $l^+ l^- b \overline b$ final state and $\cot{\theta} = 0.5$. On the right, the region where a discovery of $V_H$ to leptons or to $V h$ ($V$ is $Z$ or $W$) is possible using the ATLAS detector at LHC. The Higgs boson is assumed to decay into either $b \overline b$ or $\gamma \gamma$. The discovery region corresponds to significance larger than 5 for an integrated luminosity of 300 $fb^{-1}$.\label{fig:gauge}}
\end{figure}

The best way to search for the new gauge bosons is via their decays into electrons. The decays $Z_H \to e^+ e^-$ and $W_H \to e \nu_e$ provide the best signature at LHC, since the invariant mass resolution does not degrade for electrons with increasing mass (contrary to the muon case). The background, mainly Drell-Yan pairs, is much smaller than the signal.

The observation of the decay $Z_H \to Zh$ is an essential test for the Little Higgs model. The decay of a 120 GeV Higgs to $b \overline b$ has been studied. The final state in this mode consists of a pair of $b$-jets and a pair of leptons from the $Z$ decay. The main background is $Z$ production associated with jets. As the mass of $Z_H$ increases, the two $b$-jets from the $h$ decay merge into a double $b$-quark jet with very high $p_T$ and $b$-tagging at high $p_T$ is therefore very important. Results from the \emph{full} simulation show that it would be possible to tag these double $b$-jets with 40\% efficiency~\cite{phenomenology}. Figure~\ref{fig:gauge} (left) shows the reconstructed invariant mass of a 2 TeV $Z_H$ for an integrated luminosity of 300 $fb^{-1}$ and $\cot{\theta} = 0.5$.

The result of the combined searches for the heavy gauge bosons $Z_H$ and $W_H$ is shown in Figure~\ref{fig:gauge} (right). This figure shows the region in the $M-\cot{\theta}$ plane where the significance is larger than 5. The result using the decay $h \to \gamma \gamma$ instead of $h \to b \overline b$ is also shown in the figure. Although the $h \to \gamma \gamma$ channel has a smaller branching ratio, the good calorimetry of ATLAS allows the use of this decay as well. The signature of this channel is 2 jets from the gauge boson decay and 2 photons from the Higgs decay. Assuming that $Z_H$ and $W_H$ have the same mass, the same analysis can be used for both gauge bosons and the resulting signal is simply added. The presence of the two photons with an invariant mass equal to the Higgs mass ensures that the background is small. The background arises from either direct Higgs production or the $QCD$ production of di-photons.

\subsection{Scalar sector}
\label{sec:scalar}


New scalar states called  $\phi^0,~\phi^+,\mbox{ and }\phi^{++}$ are predicted with a large mass. These Goldstone bosons result from the breaking of the global symmetry of the model. They acquire mass via radiative corrections. At leading order all physical states are degenerate in mass. The various constraints within the model yield the following mass limits:

\begin{equation}
\frac{\sqrt{2} m_H}{v} f  ~\leq ~M_\phi ~\leq ~10~TeV
\label{eq:phi_bounds}
\end{equation}

In the scalar sector the most interesting state is the double charged particle $\phi^{++}$. If the coupling is not too small and the mass not too heavy, it may be accessible at LHC, and may provide a good test of this model. The $\phi^{++}$ has a peculiar signature since it is a double charged particle. The decay $\phi^{++} \to W^+ W^+$ is dominant ($\sim 100$ \%) and should provide the best signature at LHC.

\begin{figure}[h]
\begin{center}
    \psfig{figure=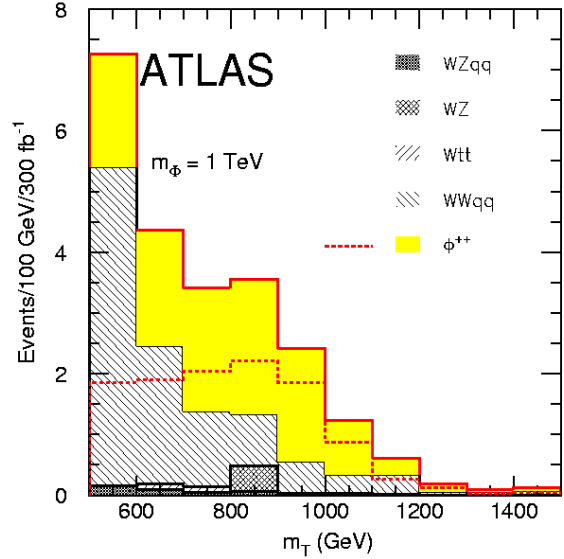,width=0.48\linewidth}
\end{center}
\caption{The mass distribution $M_T$ for a $\phi^{++}$ with a mass of 1000 GeV and $v'$ = 25 GeV. The dashed histogram represents the signal and the solid line the sum of signal and background.\label{fig:scalar}}
\end{figure}

The coupling of $\phi^{++}$ to $W^+W^+$ allows the $\phi^{++}$ production via $WW$ fusion ($qq \to qq \phi^{++} \to qqW^+W^+$). The coupling is determined by $v'$, the \emph{vacuum expectation value} of the scalar sector, which should not be too large to preserve the $SU(2)$ custodial symmetry. This decay leads to two leptons of the same charge, missing energy and secondary jets. The requirements of jets at large rapidity and high $p_T$ leptons help to decrease the background, but the process is very demanding on luminosity and the number of expected events is very low for the expected value of $v'$ (around $25~GeV$). Figure~\ref{fig:scalar} shows the background and signal for the case of $\phi^{++}$ with a mass of 1 TeV. The mass cannot be reconstructed due to the missing energy. A value of $M_{T}^{2}$ defined as $M_{T}^{2} = [E_1 + E_2 + E_T(miss)]^2 - [\vec{~p_1} + \vec{~p_2} + \vec{~~p_T}(miss)]^2$ is used instead.

\section{Conclusions}
\label{sec:conclusions}

An overview of the ATLAS capabilities to discover some of the particles predicted by the Little Higgs models has been presented. The following particles seem within the reach of LHC: the quark $T$ and the gauge bosons $Z_H$ and $W_H$. The $T$ quark could be observable up to masses of 2.5 TeV. On the other hand, the gauge bosons are accessible using their leptonic decays up to masses of 6 TeV. Furthermore, the decay of the gauge bosons to a SM gauge boson associated with a Higgs boson could provide an important test the model for masses below 2.5 TeV. The search for particles within the scalar sector is not so favorable, and in particular the $\phi^{++}$ could be discovered only if the $v'$ parameter is large enough.

\section*{Acknowledgments}
This work is the result of the combined effort of a group of people. I want to thank in particular I.~Hinchliffe, F.~Gianotti, G.~Polesello and E.~Ros. I would also like to thank the other members of the ATLAS Little Higgs Task Force: G.~Azuelos,  K.~Benslama, D.~Costanzo, G.~Couture, N.~Kanaya, M.~Lechowski, R.~Mehdiyev and D.~Rousseau.

\end{document}